\documentclass[pre,aps,twocolumn,showpacs]{revtex4-1}
\usepackage{graphicx,epsfig,amsmath,amsfonts,bm}

\def\ksi{\xi}
\def\vec#1{{\bf #1}}
\def\text#1{{\mathrm{#1}}}
\def\trace{{\mathrm{Tr}}}
\def\iff{{\rm ~if~~}}
\def\vol{\mathrm{vol}\,}

\begin{document}

\title{Radius of curvature approach to the Kolmogorov-Sinai entropy of dilute hard particles in equilibrium}
\author{Astrid S. de Wijn}
\email{A.S.deWijn@science.ru.nl}
\affiliation{Institute for Molecules and Materials, Faculty of Science, Radboud University Nijmegen, Heyendaalseweg 135, 6525~AJ Nijmegen, The Netherlands
}
\author{Henk van Beijeren}
\email{H.vanBeijeren@uu.nl}
\affiliation{Institute for Theoretical Physics, Utrecht University, Leuvenlaan 4, 3584 CE, Utrecht, The Netherlands}
\begin{abstract}
\noindent
We consider the Kolmogorov-Sinai entropy for dilute gases of $N$ hard disks or spheres.
This can be expanded in density as
$h_{\mathrm{KS}} \propto n N [\ln n a^d+ B + O(n a^d)+O(1/N)]$, with $a$ the diameter of the sphere or disk, $n$ the density, and $d$ the dimensionality of the system.
We estimate the
constant $B$
by solving a linear differential
equation for the approximate distribution of eigenvalues of the
inverse 
radius of curvature tensor.
We 
compare the resulting values of $B$ 
both to previous estimates and to existing simulation results, finding very good agreement with the latter.
Also, we compare
the distribution of eigenvalues of the inverse radius of curvature tensor resulting from our calculations to new simulation results.
For most of the spectrum the agreement between our calculations and the simulations again is 
very good.
\end{abstract}

\date{\today}

\maketitle

\section{Introduction}
It is generally believed that the approach to equilibrium of a typical many-particle system, such as a gas or liquid will depend on its dynamical properties.
Specifically, the more chaotic the system, the more rapidly
its approach to (at least local) equilibrium will proceed.
Furthermore, the apparent randomness of these systems, in spite of their fully deterministic microscopic behavior (at least for classical systems) has also been attributed to the chaotic nature of their dynamics.
Discussions of this can be found e.g.\ in books by Dorfman~\cite{bobsboek} and Gaspard~\cite{gaspbook} and review papers by Van Zon et al.~\cite{leiden,VZSzasz}.
A very common and generally used measure of chaos is the {\em Kolmogorov-Sinai entropy}, which we will denote by $h_{\mathrm{KS}}$.
In systems which are closed, the Kolmogorov-Sinai entropy $h_{\mathrm{KS}}$ equals the sum of all positive {\em Lyapunov exponents}, the average rates over very long times
of divergence (or convergence) of infinitesimal perturbations.
It describes the rate at which the system produces information about 
its phase-space trajectories, or equivalently about the distribution of density over phase space in some ensemble.
In systems with escape, the Kolmogorov-Sinai entropy has also been connected to transport coefficients~\cite{transport1,transport2,transport3,henkenbob1}.
In such systems it is no longer equal to the sum of all positive Lyapunov exponents.

Chaotic properties
such as the Lyapunov spectrum
of systems of low~\cite{henkenbob1,3d,3dposch,lagedichtheid} as well as high~\cite{onslorentz} dimensionality,
such as moving hard spheres or disks, have been studied
frequently.
Extensive simulation work has been done on their Lyapunov spectra~\cite{posch1,forster,christina}, and for
low densities analytic calculations have been done for the largest Lyapunov exponent~\cite{prlramses,ramses,leiden,jstatph}, the Kolmogorov-Sinai entropy~\cite{prlramses,lagedichtheid} and for the smallest positive Lyapunov exponents~\cite{mareschal,onszelf}.
Analytic methods employing kinetic theory have been applied to calculate chaotic properties.
Agreement between analytic calculations and numerical results is generally good, but with respect to the KS-entropy there is one notorious exception, which
is the central issue of the present paper.

In this paper we consider a system consisting of $N$ hard, spherical particles, of diameter $a$, at small number density $n$, in $d$ dimensions ($d=2,3$).
We calculate the Kolmogorov-Sinai entropy in the low density approximation, where it is expected to behave as~\cite{lagedichtheid}
\begin{align}
h_{\mathrm{KS}} = N \bar\nu A \left[-
\ln (n a^d) + B + O(na^d) + O\left(\frac{1}{N}\right)\right]~.
\label{eq:hKS}
\end{align}
The constant $A$ has been calculated by 
Van Beijeren et al.~in~\cite{lagedichtheid}, but the results found
there for $B$ were unsatisfactory.

In this paper we present a more successful calculation of $B$, through the distribution of eigenvalues of the inverse radius of curvature tensor.
The calculation presented here differs from that presented by
De Wijn in Ref.~\cite{ksentropie}, in that it is far more elegant and
less cumbersome and
the agreement of the results with values found in simulations is 
better.
On the other hand, the calculation here is less systematic and it is not
clear how to apply the results of this paper to calculating specific Lyapunov exponents, as can be done~\cite{cylinders}
with the results of Ref.~\cite{ksentropie}.

The paper is organized as follows: In section \ref{sec:lyap} we introduce Lyapunov exponents and review the properties of hard sphere dynamics in tangent space (the space in which the dynamics is described of infinitesimal deviations between nearby trajectories in phase space). In section \ref{sec:ROC} we introduce the radius of curvature tensor and its inverse, relate the KS-entropy to the time average of the trace of the inverse radius of curvature tensor and investigate the dynamics of these tensors both during free flight and at 
collisions. In \ref{sec:ROCev} we present two approximate calculations of the average distribution of the eigenvalues of the inverse radius of curvature tensor.
In section \ref{sec:discussion} we compare the results of these calculations to those of numerical simulations and we also compare the resulting value for the coefficient $B$ in Eq.\ (\ref{eq:hKS}) to those obtained in simulations and in previous calculations. Finally, in section \ref{sec:conc} we present our conclusions.

\section{\label{sec:lyap}Lyapunov exponents and dynamics of hard spheres in tangent space\label{sec:spheresdyn}}

This section is an abbreviated version of similar sections in Refs.~\cite{onszelf,ksentropie}.
It appears here to make this paper more self-contained. For more details the reader may also consult Ref.~\cite{forster}.
Consider a system with an $\cal N$-dimensional phase space $\Gamma$.
At time $t=0$ the system is at an initial point ${\vec\gamma}_0$ in this space.
It evolves with time, according to ${\vec\gamma}({\vec\gamma}_0,t)$.
If the initial conditions are perturbed infinitesimally, by $\delta{\vec\gamma}_0$, the system evolves along an
infinitesimally different path $\gamma +
\delta \gamma$, which can be specified by
\begin{align}
{\delta{\vec\gamma}({\vec\gamma}_0,t)} \label{eq:M}= {{\sf M}_{{\vec\gamma}_0}(t)\cdot \delta{\vec\gamma}_0~,}
\end{align}
with the matrix $ {\sf M}_{{\vec\gamma}_0}(t)$ defined by
\begin{align}
\label{eq:tang} {\sf M}_{{\vec\gamma}_0}(t)=\frac{d {\vec\gamma}({\vec\gamma}_0,t)}{d {\vec\gamma}_0}~.
\end{align}
The Lyapunov exponents are the possible average rates of growth
or shrinkage of such perturbations, i.e.,
\begin{equation}
\lambda_i = \lim_{t\rightarrow\infty} \frac{1}{t} \ln
|\mu_i(t)|~,
\end{equation}
where $\mu_i(t)$ is the $i$-th eigenvalue of ${\sf M}_{{\vec\gamma}_0}(t)$.
For ergodic systems,
the Lyapunov exponents are
expected to be
the same for almost all initial conditions.
For each exponent there is a corresponding eigenvector of ${\sf M}_{{\vec\gamma}_0}(t)$.

For a classical system of hard spheres without internal degrees of freedom, the phase space and tangent space may be
represented by the positions and velocities of all particles and their infinitesimal deviations,
\begin{align}
\gamma_i & = (\vec{r}_i, \vec{v}_i)~,\\
\delta\gamma_i & =  ({\vec{ \delta r}}_i, {\vec{ \delta v}}_i)~,
\end{align}
where $i$ runs over all particles and $\gamma_i$ and $\delta\gamma_i$ are the contributions of particle $i$ to $\gamma$ and $\delta\gamma$.

In the case of a purely Hamiltonian system, such as the one under consideration here, hard spheres with only the hard particle interaction, the dynamics of the system are completely invariant under time reversal.
Together with Liouville's theorem, which states that phase space volumes are invariant under the flow, this leads to the conjugate pairing rule~\cite{pairing1,pairing2},
i.e.\ for every positive Lyapunov exponent
there is a negative exponent of equal absolute value.
In systems which are
time reversal invariant, but
do not satisfy Liouville's theorem, the conditions for and the form of the conjugate pairing rule are somewhat
different~\cite{ramses}.

\begin{figure}
\includegraphics[width=6cm]{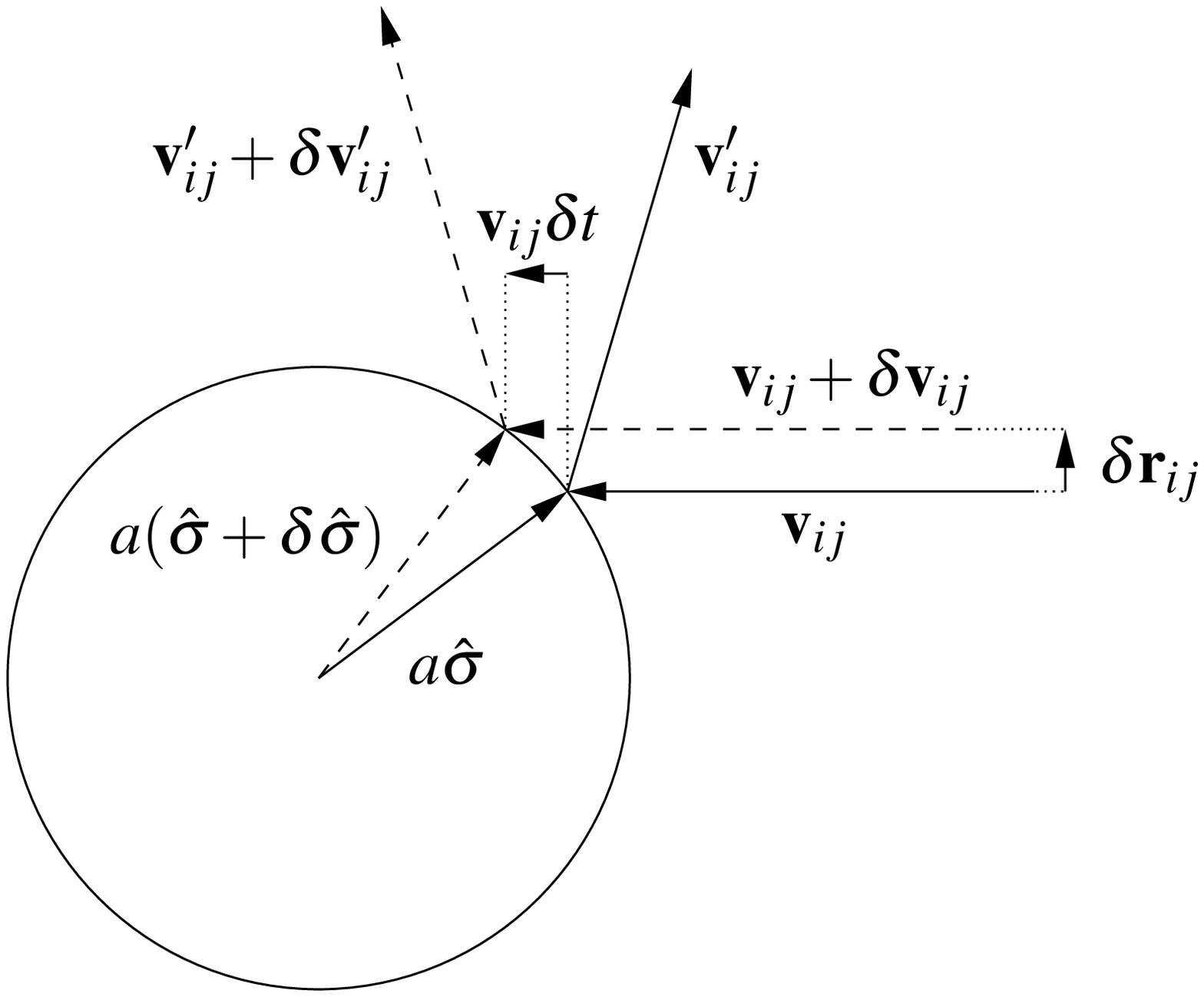}
\caption{\label{fig:bolletje} Two particles at a collision in relative coordinates.  The collision normal $\hat{\boldsymbol{\sigma}}$ is the unit vector pointing from the center of one particle to the center of the other.}
\end{figure}

The system under consideration here has only hard-core interactions.
Consequently, the evolution in phase space consists of an alternating sequence of free flights and collisions.

During free flights the particles do not interact and the positions 
change linearly with the velocities.
The components of the tangent-space vector
accordingly transform to
\begin{align}
\left(
\begin{array}{c}
\delta\vec{r}_i'\\
\delta\vec{v}_i'
\end{array}
\right)
&=
\left(\begin{array}{cc}
\extracolsep{1mm}
\rule{0mm}{0mm}{\bf 1}&(t-t_0){\bf 1}\\
\rule{0mm}{5mm}0&{\bf 1}
\end{array}\right)
\cdot \left(
\begin{array}{c}
\delta\vec{r}_{i}\\
\delta\vec{v}_{i}
\end{array}
\right)
~,
\label{eq:flight}
\end{align}
in which ${\bf 1}$ is the $d\times d$ identity matrix.

At a collision
between 
particles $i$ and $j$ momentum is exchanged between the colliding particles along the collision normal,
$\hat{\boldsymbol{\sigma}} = (\vec{r}_i-\vec{r}_j)/{a}$,
as shown in Fig.~{\ref{fig:bolletje}.
The other particles do not interact.
For convenience we switch to relative and center of mass coordinates, $\delta\vec{r}_{ij} = \delta\vec{r}_i -\delta\vec{r}_j, \delta\vec{R}_{ij} = (\delta\vec{r}_i +\delta\vec{r}_j)/2, \delta\vec{v}_{ij} = \delta\vec{v}_i -\delta\vec{v}_j$, and $\delta\vec{V}_{ij} = (\delta\vec{v}_i +\delta\vec{v}_j)/2$.
We find~\cite{prlramses,soft2}
\begin{align}
\label{eq:reldyn1} \delta\vec{r}'_{ij} &= \delta\vec{r}_{ij} - 2 {\sf S} \cdot \delta\vec{r}_{ij}~,\\ \delta\vec{R}'_{ij} &= \delta\vec{R}_{ij}~,\\ \delta\vec{v}'_{ij} &= \delta\vec{v}_{ij} - 2 {\sf S} \cdot \delta\vec{v}_{ij} - 2 {\sf Q} \cdot \delta\vec{r}_{ij}~,\\ \delta\vec{V}'_{ij} &= \delta\vec{V}_{ij}~, \label{eq:reldyn4}
\end{align}
in which ${\sf S}$ and ${\sf Q}$ are the $d \times d$ matrices 
\begin{align} \label{eq:S}{\sf S} & =
\hat{\boldsymbol{\sigma}} \hat{\boldsymbol{\sigma}}~,\\ \label{eq:Q}{\sf Q} & =
\frac{\left[(\hat{\boldsymbol{\sigma}}\cdot\vec{v}_{ij})\,{\bf 1}+\hat{\boldsymbol{\sigma}}\vec{v}_{ij}\right]\cdot \left[ (\hat{\boldsymbol{\sigma}}\cdot\vec{v}_{ij})\,{\bf 1}-\vec{v}_{ij} \hat{\boldsymbol{\sigma}}
\right]} {a (\hat{\boldsymbol{\sigma}}\cdot\vec{v}_{ij})}~,
\end{align}
where $\vec{v}_{ij}=\vec{v}_i-\vec{v}_j$.
Here the notation $\vec{a}\vec{b}$ denotes the standard tensor product of vectors $\vec{a}$ and $\vec{b}$. Note that ${\sf Q}$ transforms
vectors that are orthogonal to $\vec{v}_{ij}$ into vectors that are orthogonal to $\vec{v}'_{ij}$.
The vector ${\vec{v}}_{ij}$ is a right zero eigenvector of ${\sf Q}$, and ${\vec{v}}'_{ij}$ a left zero eigenvector.
Equations~(\ref{eq:flight}) through~(\ref{eq:Q}) determine ${\sf M}_{\gamma_0}(t)$.

\section{Radius of curvature tensor\label{sec:ROC}}

Of particular interest for the Kolmogorov-Sinai entropy is the radius of curvature tensor, as 
the former is equal to the
time average of the trace of 
the latter's inverse.
Let $\delta\vec{r}$ and $\delta\vec{v}$
represent the full $dN$ dimensional position
respectively velocity perturbations of all particles.
The {\em inverse radius of curvature tensor}, ${\cal T}(t)$ is defined~\cite{soft2,sinairoc,henkenbob1,VZSzasz,3d} as the inverse  of the radius of curvature matrix~\footnote{As was noted already in Ref.~\cite{VZSzasz} the radius of curvature tensor has the dimension of time, rather than length. However, it can be given the latter dimension by multiplying it by $V\equiv\sqrt{\sum_i|v_i|^2}$, which in the absence of an external potential is constant for hard spheres or disks.}.
It satisfies
\begin{align}
\delta\vec{v(t)} = {\mathcal T(t)}\cdot \delta\vec{r(t)}~,
\label{eq:T}
\end{align}
resulting from 
some, almost arbitrary initial ${\mathcal T(0)}$
\footnote{It can be shown that for almost all choices of ${\mathcal T}(0)$, ${\mathcal T}(t)$ rapidly becomes independent of ${\mathcal T}(0)$ with increasing $t$.}.

\subsection{KS-entropy and inverse ROC-tensor}

To  establish the relationship between the KS entropy and the trace of the inverse radius of curvature tensor we consider the time evolution over long times of the projection onto $\delta\vec{r}$ space of the infinitesimal volume evolving from an initial volume in tangent space spanned by $\delta\vec{r}(0)$ and $\delta\vec{v}(0)$.
Since the projections of all Lyapunov eigenvectors with positive exponents onto 
$\delta\vec{r}$ space are linearly independent, 
the size of this projected volume will grow roughly as the exponent of the sum of the positive Lyapunov exponents. A more precise statement is
\begin{align}
\lim_{t\to\infty}\frac{1}{t}{\ln\frac{\vol\delta\vec{r}(t)}{\vol\delta\vec{r}(0)}}=\sum_{\lambda_i>0}\lambda_i~.
\label{eq:groei}
\end{align}
Here $\vol\delta\vec{r(t)}$ denotes the volume of the projection onto $\delta\vec{r}$ space of the evolving infinitesimal volume in tangent space.
Additionally, we have the identity
\begin{align}
\delta\vec{v(t)} = \frac {d\delta\vec{r(t)}}{d t}~,
\label{eq:dvdt}
\end{align}
which holds, except at the instants of collisions $t_i$, when $\delta\vec{r}$ is reflected in a volume-preserving unitary transformation ${\mathcal U}_i$.
Together with Eq.~(\ref{eq:T}),
one obtains
\begin{align}
\vol \delta\vec{r(t)}&=
\vol \left( \prod_{i} {\mathcal U}_i
\lim_{\Delta t \rightarrow 0}
\prod^{t_i/\Delta t-1}_{n=t_{i-1}/\Delta t}(1+\mathcal T(n\,\Delta t) \Delta t)\delta\vec{r(0)}\right)\nonumber\\
&=
\prod_i \det{\mathcal U}_i
\lim_{\Delta t \rightarrow 0}
\prod_{n=t_{i-1}/\Delta t}^{t_i/\Delta t-1}\det(1+\mathcal T(n\,\Delta t)\Delta t)~,
\label{eq:det}
\end{align}
where the first product is over the sequence of all collisions and we employ the convention $\prod_{i=1}^n a_i\equiv a_n\cdots a_1$.
This leads to
\begin{align}
\lim_{t\to\infty}\frac{1}{t}{\log\frac{\vol\delta\vec{r(t)}}{\vol\delta\vec{r(0)}}}=\langle\trace\mathcal T\rangle.
\label{eq:trace}
\end{align}
By comparing Eqs.~(\ref{eq:groei}) and~(\ref{eq:trace}) one obtains the desired identity
\footnote{Although this equation can easily be obtained from the literature~\cite{VZSzasz,ksentropie}, we are not aware of it being presented in this particular form anywhere.}
\begin{align}
h_{\mathrm KS} = \langle \trace{\mathcal T} \rangle~.
\label{eq:gem}
\end{align}
Within the framework of this article, this identity is very central.
It relates the 
trace of the inverse radius of curvature tensor directly to the Kolmogorov-Sinai entropy, which we wish to calculate.

The rest of this paper is therefore dedicated to the description of the eigenvalues of the
inverse radius of curvature tensor.
We will consider their dynamics and derive approximate equations for the time evolution of their distribution.
Using these we will calculate the changes in the distribution of the eigenvalues due to both free streaming and collisions.
From this we obtain approximations for the stationary distribution of eigenvalues.

This calculation is simplified appreciably by restriction to low densities.
We may assume then that in each collision the elements of the precollisional inverse ROC tensor involving either of the colliding particles are small, due to their decrease during the free flight preceding the collision.
This assumption is violated only in collisions where one of the colliding particles had collided shortly before.
The fraction of all collisions where this is the case decreases linearly with density.

In the next section two simple approximation schemes will be presented, based on the dynamics describing
the changes in the distribution of eigenvalues of the 
inverse ROC tensor
resulting from collisions.

\subsection{Dynamics of the inverse radius of curvature tensor}

At a given collision,
between particles labelled $i$ and $j$, let ${\mathcal S}$ and ${\mathcal Q}$ be the $dN \times dN$ dimensional matrices which perform the transformations of $2{\sf S}$ and $-2{\sf Q}$
on the
relative components
in tangent space of the colliding particles
and act as zero on all other
independent
components of $\delta{\vec{v}}$ or $\delta{\vec{r}}$,
as
is described in Eqs.~(\ref{eq:reldyn1}--\ref{eq:reldyn4}).
Note that ${\mathcal Q}\cdot({\mathcal I} -{\mathcal S})$, where ${\mathcal I}$ is the $dN \times dN$ identity matrix, has $d-1$ nonzero eigenvalues and is symmetric.
One of the non-zero eigenvalues of ${\mathcal Q}\cdot{({\mathcal I}
-{\mathcal S})} $
is equal to
\begin{align}
\ksi_0 =
- \frac{2 v_{ij}
}{a \hat{\vec{v}}_{ij}\cdot \hat{\boldsymbol{\sigma}}}~,
\label{eq:eigenvalue1}
\end{align}
with $\hat{\vec{v}}_{ij}$ the unit vector in the direction of ${\vec{v}}_{ij}$. 
The corresponding eigenvector has components in the subspaces belonging to the colliding particles of ${\vec{e}}_i=-{\vec{e}}_j=\hat{\boldsymbol{\sigma}}_{ij}-(\hat{\boldsymbol{\sigma}}_{ij}\cdot
\hat{\vec{v}}'_{ij})\hat{\vec{v}}'_{ij}$ and ${\vec{e}}_k=0$ for $k\ne i,j$ along the directions of the other particles.
For $d>2$ the other $d-2$
non-zero eigenvalues are given by
\begin{align}
\label{eq:eigenvalue2}
\ksi_0 =
- \frac{2 \vec{v}_{ij} \cdot \hat{\boldsymbol{\sigma}}}{a}~,
\end{align}
with eigenvector with components ${\vec{e}}_i=-{\vec{e}}_j$ normal to both $\hat{\vec{v}}_{ij}$ and $\hat{\boldsymbol{\sigma}}$
and again ${\vec{e}}_k=0$.
The dynamics of the
inverse radius of curvature tensor at a collision can be derived by expressing $\delta\vec{v}'$ in
terms of $\delta\vec{r}'$,
\begin{align}
\delta\vec{v}' &= ({\mathcal I}-{\mathcal S}) \cdot \delta\vec{v} 
+{\mathcal Q} \cdot \delta\vec{r}\\
& = [ ({\mathcal I}-{\mathcal S}) \cdot {\mathcal T} \cdot ({\mathcal I}-{\mathcal S})^{-1} 
+{\mathcal Q}\cdot
({\mathcal I}-{\mathcal S})^{-1}] \delta\vec{r}'~.
\label{eq:evocol}
\end{align}
With $({\mathcal I}-{\mathcal S})^{-1} = ({\mathcal I}-{\mathcal S})$, we find for the
inverse radius of curvature tensor
after the collision,
\begin{align}
 {\mathcal T}'=  ({\mathcal I}-{\mathcal S}) \cdot {\mathcal T} \cdot ({\mathcal I}-{\mathcal S})
+ {\mathcal Q}\cdot ({\mathcal I}-{\mathcal S})~.
\label{eq:evofree}
\end{align}

The dynamics during free flight
follow from (\ref{eq:T}) as
\begin{align}
{\mathcal T}(t+dt) =
[
{\mathcal T}(t)
^{-1} + {\mathcal I} dt]^{-1}
\label{eq:ff}
\end{align}
The eigenvectors of ${\mathcal T}$ do not change during a free flight,
so its time evolution
may be specified by the evolution of
its eigenvalues ${\ksi}$.
From Eq.~(\ref{eq:ff})
one finds these satisfy
\begin{align}
\ksi'(t) = - \ksi(t)^2~.
\label{eq:freedrift}
\end{align}
with solution
\begin{align}
\ksi(t) = \frac 1{\ksi^{-1}(0)+t}~.
\end{align}
Note that this may be simplified by considering the dynamics of the 
radius of curvature
tensor, ${\mathcal T}^{-1}$.
For this operator, the drift velocity becomes a constant, $1$, irrespective of the eigenvalue
and the choice of time unit.
However, in this case Eq.~(\ref{eq:evocol}) becomes more complicated.

\subsection{Eigenvalue distribution at low densities}

At low densities the mean free-flight time
is given by 
\begin{align}
\bar\tau \equiv 1/\bar\nu=\frac{a\Gamma(\frac d 2)}{v_0 2 n^*\pi^{\frac{d-1}2}},
\label{eq:meanfreetime}
\end{align}
with $v_0=(k_B T/m)^{1/2}$ the thermal velocity and $n^* = n a^d$.
The probability of a particle colliding
within a free flight
time of order $a/v_0$ (this is the typical order of $
\ksi^{-1}(0)$) is of order $n^*$ and these events,
to first approximation, may be neglected.

Let us
denote a spanning set of the subspace
spanned by the eigenvectors
with non-zero eigenvalues of
$  {\mathcal Q}\cdot({\mathcal I} - {\mathcal S})$ for a specific collision
as $
{\bm \epsilon}_1$ through ${\bm \epsilon}_{d-1}$. Because ${\mathcal Q}\cdot({\mathcal I} - {\mathcal S})$ is symmetric, these vectors are both right and left eigenvectors.
The corresponding eigenvalues to linear order in $n^*$ are given by
\begin{align}
\ksi &= {\bm \epsilon}_i \cdot [({\mathcal I}- {\mathcal S})\cdot{\mathcal T}\cdot({\mathcal I}- {\mathcal S}) 
+{\mathcal Q}\cdot({\mathcal I}- {\mathcal S})] \cdot {\bm \epsilon}_i~,\label{eq:curveiegenvalue}\\
&\approx{\bm \epsilon}_i \cdot [
{\mathcal Q}\cdot({\mathcal I}- {\mathcal S})] \cdot {\bm \epsilon}_i~,
\label{eq:wlown}
\end{align}
since the elements of ${\mathcal T}$ between these eigenvectors
on average are of order $n^*$ compared to the elements of ${\mathcal Q\cdot({\mathcal I}- {\mathcal S})}$.

Under the approximation of Eq.~(\ref{eq:wlown}) the vectors ${\bm \epsilon}_i$
only depend on the collision parameters $\hat{\bm v}_{ij}$ and $\hat{\boldsymbol{\sigma}}$, as 
specified below Eqs.\ (\ref{eq:eigenvalue1}) and (\ref{eq:eigenvalue2}).
The remaining eigenvalues of ${\mathcal T}'$, to leading order in
$n^*$, can be identified as the eigenvalues of the projection
${\mathcal P}{\mathcal T}{\mathcal P}$ of the matrix ${\mathcal
T}$ onto the $dN-d+1$ dimensional space orthogonal to the $d-1$
eigenvectors ${\bm \epsilon}_i$, as follows from standard perturbation theory.
Hence it follows that they are interspersed between the precollisional eigenvalues. This is worked out in Appendix A.

These eigenvalues are distributed in roughly the same way as the
eigenvalues of the full matrix  ${\mathcal
T}$~\cite{drieitalianen,mehta}.
But, as the eigenvalues of ${\mathcal P}{\mathcal T}{\mathcal P}$
lie in between
those of ${\mathcal T}$, the distribution of these eigenvalues is
slightly narrower than that of the eigenvalues of ${\mathcal T}$.
For more details on this see 
Appendix B.

In the next section, we present two approximation schemes.
In the first scheme,
the narrowing will be ignored and the approximation
will be made that the distribution of eigenvalues of
${\mathcal P}{\mathcal T}{\mathcal P}$ is the same as
that of ${\mathcal T}$.
The resulting equation for the distribution of eigenvalues can be solved analytically.
Its solution is expressed in terms of the distribution $f_0(
\ksi_0)$ of the non-zero eigenvalues of ${\mathcal Q}\cdot({\mathcal I}-{\mathcal S})$.

In the second scheme, a somewhat more refined approximation is made, in which
we assume that each eigenvector that is changed significantly by the
collision, but not created in it, can be written as a linear combination of
exactly 2 precollisional eigenvectors. Why this is an improvement will be argued in the
discussion, where we will also briefly discuss possibilities for further
improvements.

\section{
the distribution of eigenvalues of the inverse radius of
curvature tensor\label{sec:ROCev}}

\subsection{First approximation scheme}

Now that the dynamics of the eigenvalues of the inverse radius of
curvature tensor are specified, we may write down
approximate time-evolution equations for the distribution of
these eigenvalues, $f(\ksi,t)$, which is normalized to unity.
Let $f_0(\ksi)$ be the
distribution of nonzero eigenvalues of
$ {\mathcal Q}\cdot({\mathcal I}-{\mathcal S})  $, which follows from Eqs.~(\ref{eq:eigenvalue1}) and~(\ref{eq:eigenvalue2}) and the distribution of the collision parameters $\vec{v}_{ij}$ and $\hat{\boldsymbol{\sigma}}$.
We 
equally normalize it 
to unity.
The simplest approximation for the rate of change of the distribution of eigenvalues of the inverse radius of curvature tensor is the one announced at the end of the previous section:
after a collision, the new distribution of eigenvalues is the same as the old one, except for the contributions
from the
nonzero eigenvalues of
$ {\mathcal Q}\cdot({\mathcal I}-{\mathcal S})$.
Combining this with the rate of change resulting from free streaming one obtains
\begin{align}
\frac{d}{dt} f(\ksi,t) &= 
 \frac{
\bar\nu
 (d-1)}{2 
d 
}\, [f_0(\ksi) - f(\ksi,t)]
+ \frac{\partial}{\partial \ksi} [ \ksi^2 f(\ksi,t)]~,
\label{eq:diffvergl}
\end{align}
where the single-particle collision frequency is given at low density by
Eq.\ (\ref{eq:meanfreetime}).
The first term
on the right-hand side
is due to collisions.
The first part of it is the gain.
The second part is the loss. 
Its form here is based on
our approximation that the shape of the distribution of remaining
eigenvalues is not changed in a collision. The final term is due
to the drift during free flight.

For time going to infinity, the distribution of eigenvalues
becomes stationary.
In other words, the left hand side of
Eq.~(\ref{eq:diffvergl})
becomes zero. Eq.~(\ref{eq:diffvergl})
then may be rewritten in a more convenient form, as
\begin{align}
f_0(\ksi) = f_\mathrm{stat}(\ksi) - c [\ksi^2 \frac{\partial}{\partial \ksi} f_\mathrm{stat}(\ksi) + 2 
\ksi f_\mathrm{stat}(\ksi)]~,
\label{eq:vergl}
\end{align}
where
\begin{align}
c = \frac{2 
d
}{
(d-1) 
\bar\nu}~.
\end{align}
Solutions to this equation are of the form
\begin{align}
f_\mathrm{stat}(\ksi) = \int_{\ksi}^{\infty}d\ksi_0\, f_0(\ksi_0) \frac{1}{c \ksi^2}
\exp\left(\frac{\ksi-\ksi_0}{c \ksi_0 \ksi}\right)~. \label{eq:solution}
\end{align}
Notice that for $\ksi$ small, as a consequence of Eq.~(\ref{eq:eigenvalue1}), this reduces to $f_\mathrm{stat}(\ksi) = \frac{const}{c \ksi^2}
\exp\left(\frac{-1}{c \ksi}\right)$

From Eq.~(\ref{eq:gem}) it follows that the KS entropy may be
obtained 
directly from the first moment of $f_{\mathrm {stat}}(\ksi)$. From
Eq.~(\ref{eq:solution}) we find that
\begin{align}
h_{\mathrm KS} = \frac{
Nd}{c}
 \int_{
 0}^{\infty}d\ksi_0\, f_0(\ksi_0) \exp\left(\frac{1}{c \ksi_0}\right) \Gamma\left(0,\frac{1}{c
 \ksi_0}\right),
\end{align}
where $\Gamma(,)$ denotes the incomplete gamma function,
defined by
\begin{align}
\Gamma(x,y)&=\int_y^{\infty}dt\, t^{x-1}e^{-t}\nonumber\\
&=e^{-y}\int_0^{\infty}d t(t+y)^{x-1}e^{-t}
\end{align}
At low densities
the collision frequency is low, 
so that $c$ is very large and
the product $\exp\left({1}/{(c \ksi_0)}\right) \Gamma\left(0,{1}/{(c\ksi_0)}\right)$, up to corrections of $O(n)$, is equal to
$\ln(c \ksi_0)-\gamma$, where $\gamma \approx 0.577216$ is Euler's
constant.
The Kolmogorov-Sinai entropy then becomes
\begin{eqnarray}
h_{\mathrm KS} \approx \frac{N \bar\nu (d-1)}{2} \left\{\left\langle \ln 
\frac{\ksi_0}{\bar\nu}\right\rangle +
\ln\left[ \frac{2d}{(d-1)}\right] - \gamma\right\}~.
\label{eq:resultav}
\end{eqnarray}
We note that here the brackets, instead of a time average denote an
average over the 
probability distribution for the new eigenvalues at a collision, in this case $f_0$.
This can be expressed in terms of the joint probability distribution of the collision parameters as
\begin{align}
\left\langle g(\ksi_0) \right\rangle &= \int_{0}^{\infty}d\ksi_0\, f_0(\ksi_0) g(\ksi_0) \nonumber\\
&=
\sqrt\frac{\beta m}{\pi^{d-1}}\Gamma\left(\frac d 2\right)\int d\vec{v}_i d\vec{v}_j d \hat{\boldsymbol{\sigma}} 
\theta(-\hat{\vec{v}}_{ij}\cdot \hat{\boldsymbol{\sigma}}) |(\vec{v}_i - \vec{v}_j)\cdot \hat{\boldsymbol{\sigma}}|\nonumber\\
&\phantom{=\int dv} \phi_{\mathrm M}(\vec{v}_i) \phi_{\mathrm M}(\vec{v}_j)\nonumber\\
&\phantom{=\int dv}\frac{1}{d-1} \left[ g\left(\frac{
-2 v_{ij}}{a\hat{\vec{v}}_{ij}\cdot \hat{\boldsymbol{\sigma}}}\right) + (d-2) g\left(\frac{
-2
{\vec{v}}_{ij}\cdot \hat{\boldsymbol{\sigma}}}{a}\right) \right]~,
\label{eq:defaverage}
\end{align}
where Eqs.~(\ref{eq:eigenvalue1}) and~(\ref{eq:eigenvalue2}) have been substituted and $\phi_{\mathrm M}(\vec{v})$ is the Maxwell distribution,
\begin{align}
 \phi_{\text{M}}(\vec{v}) = \left(\frac{ 2 \pi k_\mathrm{B} T}{m}\right)^{-{d}/2} \exp{\left( - \frac{m  {|\vec{v}|^2}}{2 k_\mathrm{B} T}\right)}~.
\label{eq:avint}
\end{align}
The function $\theta(x)$ is the unit step function, which vanishes for $x<0$ and equals unity for $x\ge 0$.
In general, time averages of functions of $\ksi$ may be expressed as averages over $f_{stat}$. 

In Sec.~\ref{sec:discussion} the results from Eq.~(\ref{eq:resultav}) will be 
discussed and compared with results from molecular dynamics simulations.

\subsection{Second approximation scheme\label{sec:two}}

In the previous subsection, the distribution of eigenvalues after a collision was assumed to be the same as the one before, except for the
non-zero eigenvalues of
$ {\mathcal Q}\cdot({\mathcal I}-{\mathcal S})$.
In 
Appendix A it is shown that in reality these eigenvalues are determined by the equation
\begin{align}
 \sum_i \frac{c_i^2}{\ksi-\ksi_i}=0,
\label{eq:neweigenvalues}
\end{align}
at least in the case of $d=2$, when $ {\mathcal Q}\cdot({\mathcal I}-{\mathcal S})$ has a single non-zero eigenvector~\footnote{The generalization to the case of more than one non-zero eigenvector is also discussed in 
Appendix A} ${\bm \epsilon}$, which can be expressed in terms of precollisional eigenvectors $\psi_i$ of ${\mathcal T}$ with eigenvalues $\ksi_i$, as ${\bm \epsilon}=\sum_i c_i\psi_i$.
From Eq.~(\ref{eq:neweigenvalues}) one sees
that precisely one new eigenvalue $\ksi$ originates between each subsequent pair of precollisional ones.

We can divide the $c_i$ into two categories, appreciable and almost vanishing.
In a pragmatic way, this distinction can be made by considering as appreciable the set of largest $c_i^2$ that sum to all but a small fraction of unity (for instance 0.01).
The other $c_i^2$, then, are almost vanishing.
Due to the locality of the interactions, one may argue that, in the limit of a large system, the number of appreciable $c_i$ is small compared to $N$ in most cases.

For eigenvectors with very small values of $c_i$ a new eigenvalue is found very close to an old one, typically to the left respectively the right of the old one if the nearest eigenvalue with appreciable $c_i$ is to the left respectively the right of $\psi_i$.
This may be interpreted in the following way: in Eq.\ (\ref{eq:neweigenvalues}) one may ignore all eigenvectors with almost vanishing $c_i$, because their forms and eigenvalues remain essentially unchanged.
For the remaining eigenvalues one retains the property that new eigenvalues are interspersed between the old ones.
This is illustrated in figure \ref{fig:verschuiving.eigenwaarden}.
\begin{figure}
\epsfig{figure=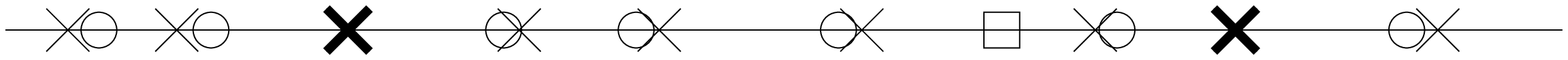,angle=270,width=8.4cm}
\caption{
An impression of eigenvalues on an interval.
The crosses denote precollisional eigenvalues, the circles and squares postcollisional ones.
The bold and the regular crosses denote eigenvalues corresponding to eigenvectors with an appreciable respectively almost vanishing $c_i^2$.
The circles and square denote new eigenvalues.
The eigenvalue indicated by the square is approximately the solution of Eq.~(\ref{eq:neweigenvalues}) with only the eigenvalues with appreciable $c_i^2$, indicated with bold crosses, contributing.
The circles almost coincide with precollisional eigenvalues with almost vanishing~$c_i^2$.}
\label{fig:verschuiving.eigenwaarden}
\end{figure}

The number of eigenvectors contributing appreciably to Eq.\ (\ref{eq:neweigenvalues}) varies from collision to collision, but it always is at least 2, because the two colliding particles cannot have collided before without intermediate collisions with other particles.
The 
actual distribution of the number of contributing eigenvectors
and the distribution of the values of the corresponding $c_i$
are not easy to determine.
In this subsection we make two simplifying assumptions:
firstly that the number of contributing eigenvectors always is just 2 and, secondly that their coefficients $c_1$ and $c_2$ are distributed isotropically, irrespective of the eigenvalues $\ksi_1$ and $\ksi_2$.
This means that these coefficients can be represented as $c_1=\cos\phi$ and $c_2=\sin\phi$, with $\phi$ distributed uniformly on the unit circle.
This assumption implies that the distribution of the corresponding original eigenvalues $\ksi_1$ and $\ksi_2$ is the same as the (as yet unknown) overall distribution of eigenvalues of the inverse radius of curvature tensor.
Obviously, these assumptions are at best approximately correct, but in our discussion we will make plausible
why, on the basis of these assumptions, a 
better approximation can be obtained for the eigenvalue distribution than the one given in Eq.\ (\ref{eq:solution}).

At a collision, the two eigenvalues $\ksi_1$ and $\ksi_2$
disappear and are replaced by a new eigenvalue, $\ksi_0$, related to $
{\mathcal Q}\cdot({\mathcal I}-{\mathcal S})$, and
the mixed eigenvalue
\begin{align}
\ksi=\ksi_1c_2^2+\ksi_2c_1^2
\label{eq:neweigenvalueinbetween},
\end{align}
as follows from Eq.\ (\ref{eq:neweigenvalues}).
The eigenvector belonging to this mixed eigenvalue is a linear combination of the two old eigenvectors,
orthogonalized to the non-zero eigenvectors
of $ {\mathcal Q}\cdot({\mathcal I}-{\mathcal S})$.
Under these approximations the collision term in Eq.~(\ref{eq:diffvergl})
is modified, leading to
\begin{eqnarray}
\frac{d}{dt} f(\ksi,t)
& =& 
\frac{
\bar\nu (d-1)}{2 d}\, [f_0(\ksi) +f_{\mathrm{coll}}[f](\ksi,t)- 2f(\ksi,t)]\nonumber\\
&\phantom{=}&+ \frac{\partial}{\partial \ksi} [\ksi^2 f(\ksi,t)]~,
\label{eq:2eigenv}
\end{eqnarray}
where $f_{\mathrm{coll}}[f](\ksi,t)$ represents the distribution of the new mixed eigenvalue after the collision, as a functional of $f(\ksi,t)$, the distribution before the collision.

Under
the assumptions described above
this distribution
can be written as
\begin{eqnarray}
f_{\mathrm{coll}}[f](\ksi,t) = \int \int d\ksi' d\ksi'' f(\ksi',t) f(\ksi'',t) h(\ksi|\ksi',\ksi'')~.
\label{eq:functional}
\end{eqnarray}
Here $h(\ksi|\ksi',\ksi'')$ is the distribution of the
new eigenvalue $\ksi$ between the eigenvalues $\ksi'$ and $\ksi''$.
The function $h$ assumes the form
\begin{eqnarray}
h(\ksi,|\ksi',\ksi'') =
\frac{1}{\pi \sqrt{(\ksi-\ksi')(\ksi''-\ksi)}}~.
\end{eqnarray}
From this an equation similar to Eq.~(\ref{eq:vergl}) can be derived.  One finds
\begin{align}
f_0(\ksi) &= 2 f_\mathrm{stat}(\ksi) -f_{\mathrm{coll}}[f_\mathrm{stat}](\ksi)\nonumber\\
&\phantom{=} - c [\ksi^2 \frac{\partial}{\partial \ksi} f_\mathrm{stat}(\ksi) + 2 \ksi f_\mathrm{stat}(\ksi)]~,
\label{eq:curvnum}
\end{align}
This equation can easily be solved numerically
and from its solution a second prediction of $h_{KS}$ can be obtained.
These results are discussed in the next section.

\section{Results and Discussion\label{sec:discussion}}

In the previous sections we have developed two closely related analytical schemes that enable us to calculate approximations for the stationary distribution of eigenvalues of the inverse radius of curvature tensor.
From this distribution we may obtain expressions for the leading order terms in the density expansion of the KS entropy of a gas of hard disks or spheres.
In particular, the 
predictions resulting from Eq.~(\ref{eq:solution}) and numerical solutions of Eq.~(\ref{eq:curvnum}) may be compared to the results of Refs.~\cite{lagedichtheid,ksentropie} and results from molecular dynamics simulations.

In Ref.~\cite{lagedichtheid}
Van Beijeren et al.\ proposed 
as approximation for the KS-entropy 
\begin{align}
h_{\mathrm KS}^{(0)} \approx \frac{N \bar\nu (d-1)}{2} \left\langle \ln \ksi_0 + \ln\left(\frac{\tau_i + \tau_j}{2} \right)\right\rangle~
,
\label{eq:resultold}
\end{align}
with $\tau_i$ the free flight time of particle $i$ since the previous collision.
Putting this in the form of Eq.~(\ref{eq:hKS}) leads
to $A = (d-1)/2$ and, after numerical integration, 
\begin{align}
B^{(0)} \approx
\begin{cases}
0.209  & \iff d=2\\
-0.583 & \iff d=3
\end{cases}
\label{eq:Bconstantelagedichtheid}
~.
\end{align}

From molecular dynamics simulations Posch and coworkers~\cite{lagedichtheid,christinapriv} found the following results for
the Kolmogorov-Sinai entropy at low densities:
\begin{widetext}
\begin{align}
h^\mathrm{num.}_{\mathrm{KS}} =
\begin{cases}
(0.499 \pm 0.001)  N\bar\nu \left(-\ln n a^d + 1.366 \pm 0.005\right) & \iff d=2\\
(1.02 \pm 0.02) N\bar\nu \left( -\ln n a^d + 0.29 \pm 0.01 \right) & \iff d=3
\end{cases}
\label{eq:ksentropiemds}
~.
\end{align}
\end{widetext}

Comparing the results of Eq.~(\ref{eq:resultav})
to those of Ref.~\cite{lagedichtheid}, Eq.~(\ref{eq:resultold}), we find that $A$ is the same, but
for $B$ one obtains corrections to Eq.~(\ref{eq:Bconstantelagedichtheid}) of the form
\begin{align}
\Delta B &= \ln\left( \frac{2d}{d-1}\right) - \gamma - \left\langle\ln\left[\frac{\bar\nu(\tau_i + \tau_j)}{2} \right]\right\rangle~.
\label{eq:corr}
\end{align}
Dorfman
et al.\ already
expected corrections
of $\ln 4
\approx 1.386$ for $d=2$
and $\ln 3\approx 1.098$ for $d=3$~\cite{logtermen},
corresponding
to the
first term
in Eq.~(\ref{eq:corr}).

In Ref.~\cite{ksentropie}, elements of the radius of curvature matrix were estimated by considering the stretching of the tangent phase space during a sequence of two collisions with free flights, it was estimated that
\begin{align}
\label{eq:oldhks}
B^\mathrm{dW} =
\begin{cases}
1.47\pm0.11
& \iff d=2\\
0.35\pm0.08
& \iff d=3
\end{cases}
\end{align}

We have evaluated the averages in
Eq.~(\ref{eq:corr})
by
integrating over the joint distribution of the collision parameters [see Eq.~(\ref{eq:defaverage})].
The values for the
parameter $B$ resulting from the first approximation scheme follow as
\begin{align}
B^{(1)} =
\begin{cases}
2 - \frac32\gamma + \ln 2 - \frac{1}{2} \ln \pi  \approx 1.255 & \iff d=2\\
\frac12 - \frac32\gamma + \ln 3 - \frac12 \ln \pi  \approx 0.160 & \iff d=3
\end{cases}
\end{align}
These results are in
reasonable agreement with the results from the molecular dynamics
simulations~\cite{christinapriv}, given in Eq.~(\ref{eq:ksentropiemds}).

More accurate results can be obtained from the second approximation scheme by numerically solving Eq.~(\ref{eq:curvnum}).
The solution for $f_\mathrm{stat}(\ksi)$ for $d=2,n = 0.001$ is displayed in Fig.~\ref{fig:curvaturedists}.
From this, one finds an additional correction to $B$ of
\begin{eqnarray}
\Delta
B = 0.086~,
\end{eqnarray}
regardless of dimensionality.
This leads to a final result for the constant $B$ of
\begin{align}
B^{(2)} \approx
\begin{cases}
1.341 & \iff d=2\\
0.247 & \iff d=3
    \end{cases}
    \label{eq:results}
\end{align}
This is in good agreement with the results from the molecular dynamics simulations, and in particular also in better agreement than the results of Ref.~\cite{ksentropie}, Eq.~(\ref{eq:oldhks}).

\subsection{Comparing approximation schemes
\label{subsec:2eigen}}

We now argue 
why the distribution of inverse radius of curvature tensor eigenvalues obtained from the two-eigenvalue approximation resulting into
Eq.\ (\ref{eq:2eigenv}) can be expected to be better than the simpler approximation, Eq.~(\ref{eq:diffvergl}), resulting
from assuming the distribution of interspersed new eigenvalues to be the same as that of the precollisional eigenvalues.
Consider
the first two moments of these distributions.
In the simple approximation these are not changed from their precollisional values.
Therefore, as mentioned already, the spectrum does not exhibit any narrowing in
collisions, as it should according to the
arguments presented before, which are supported by the calculations presented in 
Appendix B.
In the two-eigenvalue approximation, two eigenvalues $\ksi_1$ and $\ksi_2$ are sampled independently from the stationary distribution and replaced by one
interspersed eigenvalue with the value $\ksi'=\ksi_1c_2^2+\ksi_2c_1^2$, according to
Eq.\ (\ref{eq:neweigenvalueinbetween}).
The other interspersed eigenvalues
retain their
precollisional values. The coefficients $c_1$ and $c_2$ are sampled as $c_1=\cos\phi$ and $c_2=\sin\phi$, with $\phi$
distributed uniformly on the unit circle. Hence the average value of $\ksi'$ is the same as that of the precollisional
eigenvalues, as 
should be the case
(see Appendix B).
From Eq.~(\ref{eq:neweigenvalueinbetween}) and the assumed
distribution of the $c_i$ one also easily finds 
the collisional changes of the second moments.
In 
Appendix B it 
is shown that, to leading order in $1/n$ the average of the second moment
is reduced at a collision by a factor $1-(1-A)/n$, with A a constant with a value between zero and unity.
This constant, defined in the Appendix through
$
\sum_{i=1}^n\left\langle c_i^4\left(\ksi_i-\langle\ksi\rangle\right)^2\right\rangle=A/n\left\langle\sum_{i=1}^n 
\left(\ksi_i-\langle\ksi\rangle\right)^2\right\rangle$, can be calculated in the two-eigenvalue approximation
from the assumed distribution of $c_1$ and $c_2$ as
$A_2=3/4$.
Since
in most cases the new eigenvector will be composed of more than 2 precollisional eigenvectors, $A_2$ will be an upper bound
to the actual $A$.
Hence the two-eigenvalue approximation 
does lead to a narrowing of the spectrum, but 
it underestimates its extent. This is especially true in the region where $f(\ksi)$ reaches its maximum, since there the eigenfunctions of the ROC tensor tend to be carried by many particles, as one can see from Figs.\ \ref{fig:weinig-deeltjes} and \ref{fig:curvaturedists}.

The results may be improved, in principle, by considering larger sets of eigenvectors for spanning the ${\bm \epsilon}_i$.
However, to do this in a sensible way one would need the distribution of the $c_i$, preferably as a function of all $\ksi_i$.
So far no theory has been developed for this and it seems no simple task to do so.
One could of course study this distribution numerically, but that would bring one close already to a full numerical study of the
eigenvalue spectrum of the inverse radius of curvature tensor.

\begin{figure}
\epsfig{figure=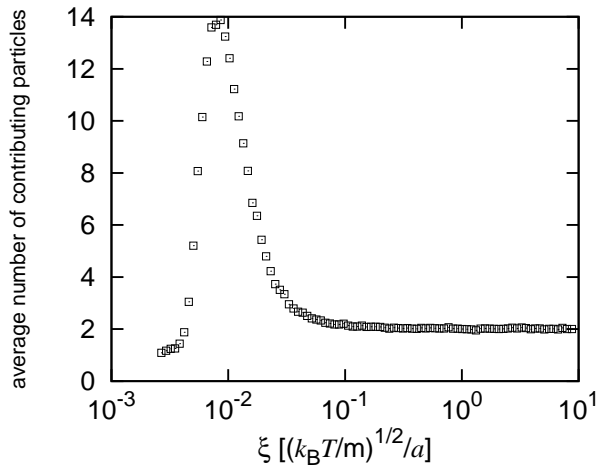,angle=270,width=8cm}
\caption{
Eigenvalue composition, as obtained from molecular dynamics simulation of 64 two-dimensional particles at $n^*$ $0.01$ in a square
box.
The figure shows an estimate for the average number of particles which contribute significantly to an eigenvector of the inverse
radius of curvature as a function of the corresponding eigenvalue $\ksi$.
This estimate is given by the ratio of the averages of the second and fourth powers of the component of the eigenvector in the
subspace of a particle.
The estimate is somewhat indirect, but it is clear that it is exact in the special case that the weight is distributed equally over a set of particles.
At this density, the collision frequency is approximately equal to $0.0354 (k_\mathrm{B} T/m)^{1/2}/a$.
In order to obtain sufficient information both at high and low densities, non-linear binning was used.
The plot only shows results for bins that contain more than 10 points.
\label{fig:weinig-deeltjes}
}
\end{figure}

In Fig.~\ref{fig:weinig-deeltjes} we plot numerical results for the average number of particles contributing to an
eigenvalue as function of $\ksi$, in a system of 64 two-dimensional particles in a square box with periodic boundary
conditions. For $\ksi$ larger than the collision frequency
most eigenvectors are carried by two particles, indicating that these are 
new eigenvectors ${\bm \epsilon}_i$.
For smaller $\ksi$ there is a rapid increase in the average number of particles carrying an eigenvector, followed by a sharp drop below
$\ksi\approx
0.25\nu \approx 0.9\times 10^{-2} (k_\mathrm{B} T/m)^{1/2}/a$.
The eigenvectors corresponding to smaller eigenvalues
have drifted for a long time, during which
in most cases the particles contributing to them have collided many times.
These eigenvectors are therefore typically carried by more particles.
But, remarkably, for very low eigenvalues the number of particles carrying the eigenvector becomes very close to 1.
This is related to the existence of particles that have not collided for several mean free times.
As a result of subsequent projections normal to new 
eigenvectors the weight of such a particle in the remaining eigenvector can increase from the original value 1/2 to values close to unity.
The contribution of these eigenvectors to the KS-entropy is very small.

We cannot directly translate the data of Fig.~\ref{fig:weinig-deeltjes} to an estimate of the number of eigenvectors contributing significantly to a newly generated eigenvector ${\bm \epsilon}_i$.
It is clear though that there is a strong correlation.
Since the new ${\bm \epsilon}_i$ are always carried by just the two colliding particles, the components along them of eigenvalues carried by several particles necessarily have to be small.
Hence, many of these are required to reconstruct any given ${\bm \epsilon}_i$.

Another, more technical approach was based on a calculation of
the distribution of elements of the 
radius of curvature~\cite{ksentropie,proefschrift}.
The results 
in 
the present paper are 
more accurate than the results of that calculation, and 
were obtained in a
more elegant way.
On the other hand, it is not directly clear to us how to further improve the accuracy of the calculation presented here,
nor if the distribution of eigenvalues of the radius of curvature could be used to calculate specific Lyapunov exponents of the system, as can be done with the distribution of the elements of the 
radius of curvature~\cite{cylinders}.

\begin{figure}[t]
\includegraphics[angle=270,width=8cm]{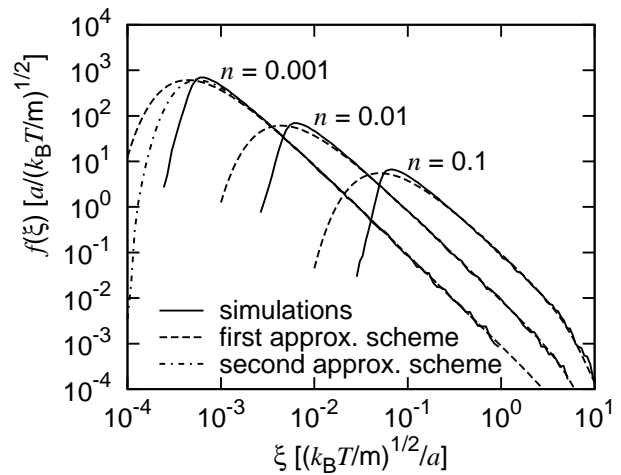}
\caption{
\label{fig:curvaturedists}
The distribution of eigenvalues of the inverse radius of curvature tensor for $d=2$ as calculated in this paper compared to results from simulations for various densities.
Dashed and dot-dashed lines 
show the results of the present calculations, while solid lines 
represent the values found in the simulations.
The results for the second approximation scheme are only shown for $n^*=0.001$, for which we calculated them with great accuracy. Similar curves would be obtained at the other densities.
The simulated systems consist of $64$ particles in square boxes
with periodic boundary conditions.  Runs were performed of 100000 collisions, and the eigenvalues of the radius of curvature were calculated and binned every 64 collisions.
In order to obtain sufficient information both at high and low densities, non-linear binning was used.
The plot only shows results for bins that contain more than 10 points.}
\end{figure}

\subsection{Distribution of eigenvalues from simulation}

In our present calculation of the Kolmogorov-Sinai entropy of
a
dilute hard-sphere gas the central quantity to be computed is the
stationary distribution of eigenvalues of the inverse radius of curvature
tensor. It is
interesting to compare the calculated distribution
to results from computer simulations.
We have performed MD simulations for a system of hard disks, in which
we calculated the radius of curvature tensor from the numerical values of $\delta\vec{r}$ and $\delta\vec{v}$ by making use of Eq.~(\ref{eq:T}) and
diagonalized it at regular time intervals. The results of these
simulations are displayed in Fig.~\ref{fig:curvaturedists}, along
with the theoretical predictions. For a
large range of eigenvalues, the calculations follow the
simulations closely,
including at very high $\xi$.
At
the lowest values of the range studied,
there appear some differences.

It can be seen from Fig.~\ref{fig:curvaturedists} that
the small eigenvalues have a 
more peaked, hence narrower
distribution than was found from the calculations.
This is due to the fact that only linear combinations of two eigenvectors were considered for ${\bm \epsilon}_i$.
In fact, as can be seen from Fig.~\ref{fig:weinig-deeltjes}
eigenvectors near the peak are generally carried by
more particles and their contribution to ${\bm \epsilon}_i$, which is carried by only two particles therefore will have a small coefficient $c_i$.
As argued above, the larger the number
of
particles carrying an eigenvector, the smaller the value of $A$
and the stronger the narrowing. And Eq.\ (\ref{eq:neweigenvalues}) reveals that for
given $\ksi$ the dynamics 
are dominated by nearby eigenvalues
and their $c_i$.
As can be seen from Fig.~\ref{fig:curvaturedists}, the distribution calculated using linear combinations of two eigenvectors
not only predicts the KS-entropy better than the approximation based on no change in the spectrum of interspersed eigenvalues, but it also follows the 
simulation results more closely for intermediate eigenvalues.

\section{Conclusions\label{sec:conc}}

In this paper we have calculated the Kolmogorov-Sinai entropy of systems consisting of hard disks or spheres
from the 
stationary distribution of the eigenvalues of the inverse radius of curvature tensor.
The dynamics of these eigenvalues consist of free streaming and collisional effects. The latter are a combination
of the generation of new eigenvalues, with a well-defined distribution, and a slight narrowing of the spectrum of
remaining eigenvalues with respect to the precollisional spectrum. Already a simple approximation,
which ignores this narrowing and assumes the spectrum to remain unchanged 
at collisions on average, reproduces the numerically
observed spectrum quite well, with a fairly accurate prediction of the KS entropy. A slightly more refined
approximation, assuming the new eigenvectors consist of just two precollisional ones, both sampled randomly
from the full distribution, does predict a narrowing of the spectrum at collisions, though it underestimates its extent.
This approximation gives quite accurate results for the KS entropy and it reproduces the
eigenvalue spectrum of the inverse radius of curvature tensor better
than the simplest approximation. The remaining underestimation of the narrowing is strongest and most clearly visible at small
eigenvalues.

In order to improve on the estimates developed here one needs more knowledge on the decomposition of the new
eigenvectors created at a collision into precollisional eigenvectors. This is highly nontrivial, however.

It should be noted that, though the
specific dynamics of the inverse radius of curvature tensor are different for
other high-dimensional systems, such as the high-dimensional Lorentz gas~\cite{onslorentz}
(which has uniformly convex scatterers), their
overall behavior is generic for all systems consisting of many particles.

An attractive approach seems describing the dynamics of the
eigenvalues of the inverse radius of curvature tensor by means of a Fokker-Planck
equation.
In order to do so one needs expressions for the local
drift and diffusion of the eigenvalues due to collisions. For this
again
more knowledge is needed of the way new eigenvectors are
composed of old ones.
Furthermore, the Fokker-Planck equation is a good approximation for systems where the dynamics consist of small jumps, while in the present case also large jumps happen.
However, these mostly will occur for large values of $\xi$, where the dynamics is dominated by the drift. Therefore the Fokker-Planck equation may still be a good approximation.

Finally, the expressions derived here for the dynamics of the inverse radius of curvature tensor, and the equations for the
distribution of its eigenvalues, Eqs.~(\ref{eq:diffvergl}) and~(\ref{eq:2eigenv}),
can also be used for systems in a
stationary non-equilibrium state.
In such a
state, the distributions of velocities and collision parameters are
different and
one has to take this into account when calculating the averages or the source terms in
Eqs.~(\ref{eq:diffvergl}) and~(\ref{eq:2eigenv}).

\begin{acknowledgments}
ASdW's work is financially supported by a Veni grant of Netherlands Organisation for Scientific Research (NWO).
HvB acknowledges support by the Humboldt Foundation.
This work was also supported by the Foundation for Fundamental Research on Matter (FOM), which is financially supported by NWO.
\end{acknowledgments}

\begin{appendix}

\section{Eigenvalues of a submatrix\label{appendix:bewijsje}}

Let ${\mathcal P}_1$ through ${\mathcal P}_{d-1}$ be the projection operators which project out the $d-1$ orthogonal unit vectors ${\bm \epsilon}_1$ through ${\bm \epsilon}_{d-1}$.
We are interested in the (distribution of) nonzero eigenvalues of submatrix ${\mathcal{P T P}}$ of a symmetric
matrix ${\mathcal T}$,  with ${\mathcal P}={\mathcal P}_1 \cdot \ldots \cdot {\mathcal P}_{d-1}$.
We shall determine these by considering the nonzero eigenvalues $\ksi_i'$ of $\tilde{\mathcal T}' = {\mathcal P}_x \tilde{\mathcal T} {\mathcal P}_x$, and their corresponding eigenvectors $\eta_i$.
The eigenvalues of $\mathcal{P T P}$ can now be determined by applying this procedure $d-1$ times
projecting subsequently normal to each of the $d-1$ non-vanishing eigenvalues of $({\mathcal I} -{\mathcal S}) \cdot {\mathcal Q}$. Thus one obtains matrices $\mathcal{\tilde{T}}$ with a decreasing number of nonzero eigenvalues.

Let $\psi_1$ through $\psi_{z}$ be the $z$ normalized eigenvectors of a matrix $\tilde{\mathcal T}$ with corresponding eigenvalues $\ksi_i$.
Let us write the unit vector that is to be projected out, ${\bm \epsilon}_x$, and an eigenvector $\eta$ of $\tilde{\mathcal T}'$ with eigenvalue $\ksi'$ in terms of the eigenvectors of $\mathcal{\tilde{T}}$,
\begin{align}
{\bm \epsilon}_x = \sum_i c_{i} \psi_i~,\\
\eta = \sum_i \beta_{i} \psi_i~,
\end{align}
with
\begin{align}
\sum_i c_{i}^2 = \sum_i \beta_{i}^2 =1~.
\label{eq:sumckwadraat}
\end{align}
As $\eta$ has no component 
along ${\bm \epsilon}_x$,
we may write
\begin{align}
\ksi' \eta
& = \ksi' \sum_i \beta_{i} \psi_i\\
& = \tilde{\mathcal T}'\eta
= {\mathcal P}_x {\mathcal{\tilde{T}}} \eta = {\mathcal P}_x \sum_i \beta_{i} \ksi_i \psi_i\\
& = \sum_i (\beta_{i}\ksi_i - \mu c_{i}) \psi_i ~,
\label{A6}
\end{align}
with $\mu$ a constant such that
\begin{align}
{\bm \epsilon}_x \cdot \ksi' \eta =  \sum_i c_{i} (\beta_{i} \ksi_i - \mu c_{i}) = 0~.
\label{eq:mu}
\end{align}
By taking the inner product of Eq.~(\ref{A6}) with a given $\psi_i$ one finds that,
\begin{align}
\beta_{i} =  - \frac{\mu c_{i}}{\ksi' -\ksi_i}~.
\label{eq:beta}
\end{align}
By substituting Eq.~(\ref{eq:beta}) into Eq.~(\ref{eq:mu}), and dividing by $\mu$ and $\ksi'$, we find that
\begin{align}
\sum_i \frac{c_i^2 }{\ksi_i - \ksi'} = 0~.
\label{eq:sum}
\end{align}
From this equation it follows directly that between each subsequent pair $\ksi_i$ and $\ksi_{i+1}$ there must be 
precisely one solution for $\ksi'$.

\section{Narrowing of eigenvalue spectrum\label{appendix:narrowing}}
In order to investigate the narrowing of the 
spectrum of eigenvalues as result of a collision we rewrite Eq.~(\ref{eq:sum}), multiplying it by $-\prod_i (\ksi'-\ksi_i)$ and find
\begin{align}
&\ksi'^{n-1}-\sum_i(1-c_i^2)\ksi_i\ksi'^{n-2}+\sum_{i< j}(1-c_i^2-c_j^2)\ksi_i\ksi_j\ksi'^{n-3}\nonumber\\
&\null+\cdots=0,
\label{eq:polynome}
\end{align}
where we have made use of Eq.~(\ref{eq:sumckwadraat}).
By comparing the coefficient of $(\ksi')^{n-2}$ to the coefficient in the eigenvalue equation for $\ksi'$,
one immediately obtains
\begin{align}
\sum_{i=1}^{n-1}\ksi'_i=\sum_{i=1}^n(1-c_i^2)\ksi_i.
\label{eq:eigenvaluesum}
\end{align}
In section \ref{sec:ROC} we have found that the the eigenvectors ${\bm \epsilon}_z$ consist of equal and opposite components on two arbitrarily determined (colliding) particles along a unit vector that is distributed isotropically in d-dimensional space.
From this it follows immediately that the average of $c_i^2$ over many collisions has to be equal to $1/n$.
If in Eq.~(\ref{eq:eigenvaluesum}) we replace $c_i^2$ by this average we find that the mean value of the interspersed
eigenvalues on average is the same as that of the precollisional ones.
Note that Eqs.~(\ref{eq:diffvergl}) and (\ref{eq:2eigenv}) both satisfy this property.

Similarly, from the coefficient of $(\ksi')^{n-2}$ one obtains the identity
\begin{align}
\sum_{i<j}^{n-1}\ksi'_i\ksi'_j=\sum_{i<j}^{n}(1-c_i^2-c_j^2)\ksi_i\ksi_j.
\label{eq:eigenvalueproductsum}
\end{align}
Combining this with Eq.~(\ref{eq:eigenvaluesum}) one finds the identity
\begin{widetext}
\begin{eqnarray}
\sum_{i=1}^{n-1}\left\langle
(\ksi'_i-\langle\ksi\rangle)^2\right\rangle&=&\frac1{n^2}
\left\langle\left(\sum_{i=1}^{n}(\ksi_i-\langle\ksi\rangle)\right)^2\right\rangle+\left(1-\frac {2
} n - \frac 1{n^2}\right)\sum_{i=1}^{n}
\langle(\ksi_i-\langle\ksi\rangle)^2\rangle+\sum_{i=1}^n \langle c_i^4(\ksi_i-\langle\ksi\rangle)^2\rangle~.
\label{eq:eigenvalueproductsum2}
\end{eqnarray}
\end{widetext}
Here the brackets indicate an average over many subsequent collisions. We used the identity
\begin{eqnarray}
\langle c_i^2c_j^2\ksi_i\ksi_j\rangle=\frac 1 {n^2}\langle\ksi_i\ksi_j\rangle\ \ \ i\ne j,
\end{eqnarray}
and we introduced the symbol $\langle\ksi\rangle$ defined by
\begin{eqnarray}
\label{eq:B2}
\langle\ksi\rangle&=&\frac 1 n\langle\sum_{i=1}^n \ksi_i\rangle,\nonumber\\
\end{eqnarray}
To leading order in $1/n$ the terms proportional to $1/n^2$ in
Eq.~(\ref{eq:eigenvalueproductsum2}) may be ignored. We introduce the constant $A$ defined through
\begin{eqnarray}
\sum_{i=1}^n\langle c_i^4(\ksi_i-\langle\ksi\rangle)^2\rangle&=&\frac A n 
\sum_{i=1}^n \langle(\ksi_i-\langle\ksi\rangle)^2\rangle~.
\label{eq:B7}
\end{eqnarray}
Note that $A$ is smaller than 1, since $c_i^4<c_i^2$ and $<c_i^2>=1/n$. To leading order in $1/n$ combination of Eqs.~(\ref{eq:eigenvalueproductsum2}) and~(\ref{eq:B7}) leads to the reduction factor mentioned in section \ref{subsec:2eigen}.
Obviously, the smaller $A$, the stronger the narrowing. This should apply also locally, implying stronger narrowing in regions where the eigenvectors tend to be carried by more particles.
\end{appendix}

\end{document}